# Quasiparticles-mediated thermal diode effect in Weyl Josephson junctions

Pritam Chatterjee [1,2,*] and Paramita Dutta [3,†]

[1] *Institute of Physics, Sachivalaya Marg, Bhubaneswar-751005, India*
[2] *Homi Bhabha National Institute, Training School Complex, Anushakti Nagar, Mumbai 400094, India*
[3] *Theoretical Physics Division, Physical Research Laboratory, Navrangpura, Ahmedabad-380009, India*

We theoretically show quasiparticles-driven thermal diode effect (TDE) in an inversion symmetry-broken (ISB) Weyl superconductor (WSC)-Weyl semimetal (WSM)-WSC Josephson junction. A Zeeman field perpendicular to the WSM region of the thermally-biased Weyl Josephson junction (WJJ) induces an asymmetry between the forward and reverse thermal currents, which is responsible for the TDE. Most interestingly, we show that the sign and magnitude of the thermal diode rectification coefficient is highly tunable by the superconducting phase difference and external Zeeman field, and also strongly depends on the junction length. The tunability of the rectification, particularly, the sign changing behavior associated with higher rectification enhances the potential of our WJJ thermal diode to use as functional switching components in thermal devices.

## I. INTRODUCTION

Diodes are two-terminal components of electronic devices, that prefer charge current to flow in one direction over the opposite direction due to the directional dependence of the resistance in the circuit [1]. Initially, diodes were fabricated using metal and crystalline minerals, followed by semiconductors [1, 2]. Recently, superconducting diodes have attracted substantial interest due to the possibility of enhanced rectification of dissipationless supercurrents [3–19], in contrast to the traditional diodes that are based on the resistive transport allowing dissipation. The seminal work by Ando *et al.* on the magnetically controllable superconducting diode effect (SDE) in an artificial superlattice made of Nb/V/Ta boosted the research in this direction [3]. It has been followed by several theoretical works and experiments to show the SDE using topological insulators, Rashba nanowire, noncentrosymmetric superconductor/ferromagnet multilayer, Rashba superconductor, quasi one-dimensional (1D) superconductor, chiral superconductor, twisted multilayer graphene, high temperature cuprate, etc. [8, 20–29].

Superconducting Josephson diode effect (JDE) has drawn particular attention of the community because of the additional freedom of the tunability of the nonreciprocal critical current by the phase difference externally [7, 28, 30–57]. Most works employed the combined effect of the inversion and time-reversal symmetry breaking resulting in finite-momentum Cooper pairs responsible for the nonreciprocity. This mechanism is very effective in enhancing the rectification, compared to the traditional noncentrosymmetric bulk materials [58–60].

In many works, topological phases of quantum matters like topological insulators, topological semimetals, Majorana bound states etc., have been extensively utilized to show the diode effects [7, 44, 50]. Very recently, SDE has been proposed in tilted WSM which is one of the recently discovered topological materials [61–66]. The low density of states around the gapless nodal points, known as Weyl nodes, helps in breaking the spatial and time-reversal symmetries [67], while the appearance of multiple Fermi pockets and pairing channels help enhance the SDE [67].

Similar to the charge current, heat current can also show nonreciprocity, giving rise to the heat or thermal diodes which are promising for thermal circuits and devices like thermal isolation, cooling, caloritronics, etc. [68–70]. It is much less explored compared to the charge counterpart. The concept of thermal diode was developed using 1D lattices, metal-dielectric interfaces, normal metal tunnel junction, quantum Hall conduc-tor, WSM, normal metal/superconductor hybrid junction [70–75]. Study of the superconducting hybrid junctions in this context is interesting since there appear some recent works on the thermal current in superconducting junctions including JJs which show distinct and interesting properties [76–88]. It is very timely and relevant to ask: is it possible to find TDE using JJ? If yes, is it possible to control the TDE externally? To the best of our knowledge, there exists only one work on the passive thermal rectification through the edge states in topological JJ based on two-dimensional topological insulator [89].

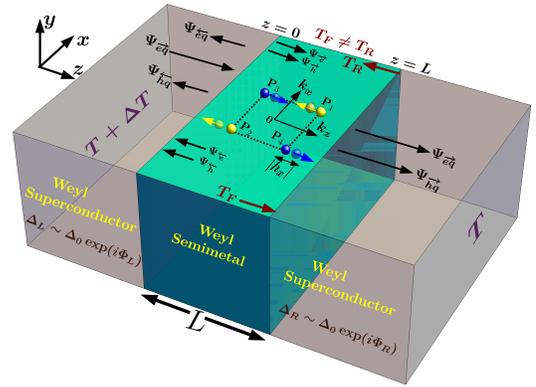

FIG. 1. Schematic of a thermally-biased inversion-symmetry broken Weyl Josephson junction. The shifting of the Weyl nodes of positive chirality (yellow balls) and the negative chirality (blue balls) in opposite directions by the external Zeeman field leads to an asymmetry between the forward and reverse quasiparticle transmission probabilities ($T_\mathrm{F} \neq T_\mathrm{R}$).

---
[*] pritam.c@iopb.res.in
[†] paramita@prl.res.in

In this work, we show TDE in three-dimensional (3D) WJJ where ISB WSM is sandwiched between two WSCs as shown in Fig. 1, with an external Zeeman field applied in the WSM region along the direction perpendicular to the junction. This generates an asymmetry between the quasiparticle transmission probabilities along the forward and reverse direction. We show that the sign and the magnitude of the thermal diode rectification coefficient in our WJJ strongly depends on the (i) length of the junction, (ii) external Zeeman field, and (iii) superconducting phase difference. Most importantly, the latter two offer external tunability to the rectification, and can boost the rectification coefficient to 90% tending to ideal diode behavior. The tunable sign changing phenomenon in our 3D thermal diode based on the topology of Weyl materials immensely enhances the importance of our work.

## II. MODEL AND FORMALISM

We consider an ISB WSM described by the Hamiltonian $\mathcal{H}_w = \sum_{\mathbf{k}} \Phi_{\mathbf{k}}^{\dagger} H(\mathbf{k}) \Phi_{\mathbf{k}}$ where

$$H(\mathbf{k}) = k_x \eta_x \sigma_z + k_y \eta_y \sigma_0 + (\kappa_0^2 - |\mathbf{k}|^2)\eta_z \sigma_0 \\ - \alpha k_y \eta_x \sigma_y + \beta \eta_y \sigma_y, \quad (1)$$

and $\Phi_{\mathbf{k}}^{\dagger} = (c_{A,\uparrow,\mathbf{k}}^{\dagger}, c_{A,\downarrow,\mathbf{k}}^{\dagger}, c_{B,\uparrow,\mathbf{k}}^{\dagger}, c_{B,\downarrow,\mathbf{k}}^{\dagger})$ [88, 90, 91]. The Pauli matrices $\eta$ and $\sigma$ act on the orbital (A,B) and spin $\uparrow(\downarrow)$ degrees of freedom, respectively. Here, $\kappa_0$, $\alpha$ and $\beta$ are model parameters. The four Weyl nodes of the ISB Weyl Hamiltonian in Eq.(1) are located at $P_{1,2} = \pm\left(\beta, 0, \sqrt{\kappa_0^2 - \beta^2}\right)$ and $P_{3,4} = \pm\left(\beta, 0, -\sqrt{\kappa_0^2 - \beta^2}\right)$ assuming $0 < \beta < \kappa_0$ where $P_1$ and $P_2$ ($P_3$ and $P_4$) carry the positive (negative) chirality to form the time-reversed pairs. Weyl nodes are situated at $k_y = 0$ for $\alpha \neq 0$ with the separation regulated by $\beta$ ($<1$).

In order to linearize the Hamiltonian in Eq.(1) around the Weyl points, we define a new set of basis: $c_{\sigma,k}^{\eta} = \frac{1}{\sqrt{2}}\left(c_{\eta\uparrow k} \pm c_{\eta\downarrow k}\right)$ where, $c_{\sigma,\vec{k}}^{\eta}$ corresponds to $\sigma$ (up/down) spin along the x-direction. The low-energy Hamiltonian in Eq.(1) can be linearized to write as a summation of four $2 \times 2$ Hamiltonians around the four Weyl nodes as $\mathcal{H}_w = \sum_{\lambda=1}^{4} \sum_{\mathbf{k}}' \chi_{\lambda,\mathbf{k}}^{\dagger} H_\lambda(\mathbf{k}) \chi_{\lambda,\mathbf{k}}$ where

$$H_{1(2)}(\mathbf{k}) = (k_x \mp \beta) \sigma_x + k_y \sigma_y + \left(k_z \mp \sqrt{\kappa_0^2 - \beta^2}\right) \sigma_z, \\ H_{3(4)}(\mathbf{k}) = (k_x \mp \beta) \sigma_x + k_y \sigma_y - \left(k_z \pm \sqrt{\kappa_0^2 - \beta^2}\right) \sigma_z, \quad (2)$$

where $k_y$ and $k_z$ are scaled by $1/\alpha$ and $1/(2\sqrt{\kappa_0^2 - \beta^2})$, respectively. The spinors for the Weyl nodes $\chi_{\lambda,\mathbf{k}}^{\dagger}$ with $\lambda \in \{1,2,3,4\}$ are given by $\chi_{1,\mathbf{k}}^{\dagger} = \chi_{3,\mathbf{k}}^{\dagger} = (c_{\uparrow,\mathbf{k}}^{(B)\dagger}, c_{\downarrow,\mathbf{k}}^{(A)\dagger})$ and $\chi_{2,\mathbf{k}}^{\dagger} = \chi_{4,\mathbf{k}}^{\dagger} = (c_{\uparrow,\mathbf{k}}^{(A)\dagger}, c_{\downarrow,\mathbf{k}}^{(B)\dagger})$.

To achieve the TDE in our ISB WJJ, we break the time-reversal symmetry by applying an external Zeeman field $\mathbf{h} = (h_x, h_y, h_z)$ perpendicular to the junction as shown in Fig. 1. The Zeeman field couples to the spin degree of freedom via $H_{\text{int}}^{\lambda} = \eta_0 \mathbf{h} \cdot \boldsymbol{\sigma}$ in the $\Phi_{\mathbf{k}}$ basis, which takes the form as $\tilde{H}_{\text{int}}^{\lambda} = h_x \sigma_z$ in the transformed basis $\chi_{\lambda,\mathbf{k}}$. Here, we conider only $h_x$ since it is the only significant term in the low-energy regime [90, 91]. It is responsible for the coupling between electrons from the same nodes and also from different nodes. The presence of this field causes shifting of the Weyl nodes by an amount $\pm h_x$ along the $k_z$ direction (see Fig. 1) which is responsible for the manifestation of the TDE.

Now, we attach two WSCs at the opposite sides of the WSM (see Fig. 1). The Bogoliubov–de Gennes (BdG) Hamiltonian for the positive chirality is given by,

$$H_{\text{BdG}} = \begin{pmatrix} \mathcal{H}'_{\text{BdG}} & O_{4 \times 4} \\ O_{4 \times 4} & \mathcal{H}''_{\text{BdG}} \end{pmatrix} \quad (3)$$

in the Nambu basis $(\chi_{1,\uparrow}, \chi_{1,\downarrow}, \chi_{2,\downarrow}^{\dagger}, -\chi_{2,\uparrow}^{\dagger}, \chi_{2,\uparrow}, \chi_{2,\downarrow}, \chi_{1,\downarrow}^{\dagger}, -\chi_{1,\uparrow}^{\dagger})$ where $O_{4\times 4}$ as the null matrix. After the Block-diagonalization, it is sufficient to focus on a single block since there are two essentially identical blocks ($\mathcal{H}'_{\text{BdG}} = \mathcal{H}''_{\text{BdG}} = \mathcal{H}_{\text{BdG}}$) in the BdG Hamiltonian given by

$$\mathcal{H}_{\text{BdG}} = h(\mathbf{r})\nu_0 \sigma_z - \mu(\mathbf{r})\nu_z \sigma_0 - i\nu_z \partial_{\mathbf{r}} \cdot \boldsymbol{\sigma} \\ + Re[\Delta_s(\mathbf{r})]\nu_x \sigma_0 - Im[\Delta_s(\mathbf{r})]\nu_y \sigma_0, \quad (4)$$

which can be found after a unitary transformation of Eq.(2). Here, the Pauli matrix $\nu$ acts on the particle-hole space, $h(\mathbf{r}) = h_x \Theta(z) + h_x \Theta(L - z)$, the pairing potential term: $\Delta_s(\mathbf{r}) = \Delta(T)e^{i\Phi_L}\Theta(-z) + \Delta(T)e^{i\Phi_R}\Theta(z - L)$, where $\Theta(z)$ is the Heaviside step function, $\Phi_L(\Phi_R)$ is the phase of the left (right) superconductor, $\phi$ is the phase difference ($\phi = \Phi_L - \Phi_R$) and $L$ is the length of the middle WSM region. A similar Hamiltonian can be written for the negative chirality Weyl nodes. The temperature dependence of the superconducting gap is taken as $\Delta(T) = \Delta_0 \tanh(1.74\sqrt{T_c/T - 1})$ with $\Delta_0$ as the gap at $T = 0$. The chemical potential term is taken as follows: $\mu(\mathbf{r}) = \mu_N \Theta(z) + \mu_N \Theta(L-z) + \mu_S \Theta(-z) + \mu_S \Theta(z-L)$ where $\mu_{N(S)}$ is the chemical potential of the semimetal (superconducting) region of our WJJ.

In order to calculate the thermal current driven by the quasiparticles with energy $\epsilon > \Delta_0$, we apply a thermal gradient across the junction by maintaining different temperatures $T + \Delta T$ and $T$ at the two superconductors. We call it positive (negative) thermal gradient when the temperature of the left (right) WSC is higher than that of right (left) WSC. Note that for $\epsilon < \Delta_0$, there are no propagating modes to carry the thermal current. The thermal current in our junction is entirely carried by the quasiparticles [76, 79, 81, 85, 86]. To calculate the forward (backward) current carried by the quasiparticles, we need to evaluate the forward (backward) transmission probability of the quasiparticles $T_F$ ($T_R$). We use the scattering matrix formalism to find the total transmission probabilities defined as $T_{F(R)} = \left|t_{ee}^{F(R)}\right|^2 + \left|t_{he}^{F(R)}\right|^2$ where $t_{\alpha\beta}^{F(R)}$ represents the forward (backward) transmission amplitude of $\alpha$-like quasiparticles as $\beta$-like quasiparticles ($\alpha, \beta \in \{e, h\}$). We numerically solve it by matching the normalized wave functions at two junctions of our model (as shown in Fig. 1). We refer to Appendix A for the details of the formalism. Note that the scattering amplitudes obey the unitarity condition for each direction as $\left|r_{ee}^F\right|^2 + \left|r_{he}^F\right|^2 + \left|t_{ee}^F\right|^2 + \left|t_{he}^F\right|^2 = 1$, where $r_{\alpha\beta}^F$ is the amplitude of the reflection of the incident $\alpha$ particle (along the forward direction) as $\beta$ particle. Similar condition holds for the reverse direction too.

To find the thermal current flowing along $z$-direction, we integrate out the parallel components of the momenta, denoted as $k_\parallel = (\epsilon + \mu_N) \sin 2\alpha_e$, and finally, define the thermal current per unit temperature gradient namely, thermal conductance (in units of $k_B^2/2\pi h$) as [88],

$$\kappa_{F(R)} = \int_0^{\frac{\pi}{4}} \int_{|\Delta(T)|}^{\infty} (\epsilon + \mu_N)^2 \sin 4\alpha_e \, T_{F(R)} \left(-\frac{\partial f}{\partial T}\right) \epsilon d\epsilon d\alpha_e \quad (5)$$

where, $f$ and $\alpha_e$ are the equilibrium Fermi distribution function and the angle of incidence at the WSM region, respectively. We refer to the Appendix A and B for other details of the model and formalism.

## III. PHASE DEPENDENT THERMAL CURRENT

We start by discussing the behaviors of the forward (when positive thermal gradient) and backward (when negative thermal gradient) thermal conductance individually and their phase dependence, sensitivity to the external Zeeman field, and junction length dependence shown in Fig. 2. In Fig. 2(a), we see that in the absence of any magnetic field, the thermal conductances along the two opposite directions, being $2\pi$-periodic, are exactly equal to each other ($\kappa_F = \kappa_R$). When we apply the Zeeman field, an asymmetry between the forward and reverse current grows and we find $\kappa_F \neq \kappa_R$ except some values of $\phi$ (see Fig. 2(b)). This nonreciprocity of the current describing the diode effect stems from the asymmetry between the transmission probabilities of the quasiparticles along the forward and reverse directions. Notably, the TDE in our junction is tunable externally since it is sensitive to the superconducting phase and Zeeman field. On further investigations on the behavior of the currents with the magnetic field, in Fig. 2(c), we observe that for the entire range of $|h_x L|$, $\kappa_F$ and $\kappa_R$ differ from each other except a few crossings where they become equal to each other. The crossings appear at $|h_x L| = n\pi/2$, where $n$ is an integer. The condition for the vanishing diode effect can be explained by using the analytical expressions mentioned in the upcoming section. In all these three figures, we present our numerical results for a particular junction size. In Fig. 2(d), we plot $\kappa$ for both directions as a function of the junction length. It is clear that in the short junction limit ($L < \xi$ with $\xi = \hbar v_F/\Delta_0$ as the superconducting coherence length), we observe a significant difference between the forward and reverse currents indicating an efficient diode effect in this regime. On the other hand, for the long junction regime ($L > \xi$), the two currents are very close to each other, leading to a vanishingly small diode effect.

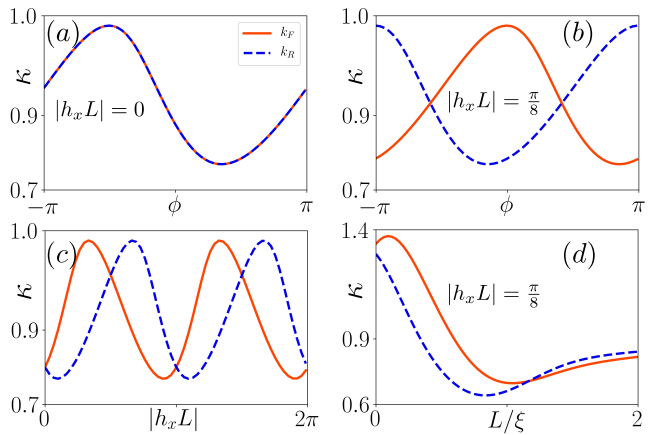

FIG. 2. Thermal conductance (in units of $k_B^2/2\pi h$) along the forward and reverse directions for (a) $|h_x L| = 0$ and (b) $|h_x L| = \pi/8$ keeping $L/\xi = 0.5$, for (c) $\phi = 0.16\pi$ keeping $L/\xi = 0.5$, and for (d) $\phi = 0.16\pi$ keeping $|h_x L| = \pi/8$. Other parameters are: $\mu_N = 0.5\Delta_0$, $\mu_S = 100\Delta_0$, and $T/T_c = 0.3$.

In order to understand the distinct behaviors of the currents, we now find the analytical expressions for the forward and the reverse transmission probability corresponding to the positive and negative thermal gradient, respectively, by setting $\tilde{\alpha}_e \approx 0$. Note that, for all nonzero incidence angles, we solve the problem numerically. For the normal incidence ($\tilde{\alpha}_e \approx 0$) and in the short junction limit ($L \ll \xi$), the expressions are given by,

$$T_{F(R)}(\epsilon, \phi) \simeq (\epsilon^2 - \Delta_0^2) / \left[\epsilon^2 - \Delta_0^2 \cos^2(\frac{\phi}{2} \mp |h_x L|)\right] \quad (6)$$

assuming $\Phi_L = -\Phi_R = -\phi/2$. It is evident from the above expression that there exists an asymmetry between the forward and backward transmission probabilities of the quasiparticles ($T_F - T_R \neq 0$) due to the presence of the term $|h_x L|$ being proportional to the external Zeeman field. This clearly explains the TDE. The transmission as well as the TDE is not sensitive to whether $h_x$ is positive or negative. Our expression for the transmission probability for the quasiparticles in the normal incidence limit ($\alpha_e, \tilde{\alpha}_e \approx 0$) as mentioned in Eq. (6) is very similar to the transmission probability derived in Ref.[92] when $|h_x L| = 0$. Note that, we can approximate the Andreev bound state energy formed in the junction for the case $\epsilon < \Delta_0$ by setting the denominator of the transmission functions to zero as $E_{F(R)}(\phi) \sim \pm\Delta_0 \cos(\frac{\phi}{2} \mp |h_xL|)$. On the other hand, in the long junction limit ($L \gg \xi$), the expression for the quasiparticles' transmission probability takes the form,

$$T_{F(R)}(\epsilon, \phi) \simeq (\epsilon^2 - \Delta_0^2) / \left[\epsilon^2 - \Delta_0^2 \cos^2(\frac{\phi}{2} + \mu_N L \mp |h_xL|)\right] \quad (7)$$

and the corresponding Andreev bound state energy for $\epsilon < \Delta_0$ can be approximated as, $E_{F(R)}(\phi) \sim \pm\Delta_0 \cos(\frac{\phi}{2} + \mu_N L \mp |h_xL|)$. Thus, it is evident that an asymmetry appears between the forward and reverse transmission probabilities of the quasiparticles across the junction in the long junction limit too, clearly indicating the TDE. However, the TDE in our junction is suppressed because of the large $\mu_N$ value. The term proportional to the magnetic field is very small compared to the chemical potential. Note that, the scattering formalism used in the present manuscript does not include the contributions for $\epsilon < \Delta_0$. Thus, our numerical calculations hold for only quasiparticles with $\epsilon > \Delta_0$. We refer to Appendix C for some additional results of the transmission probability.

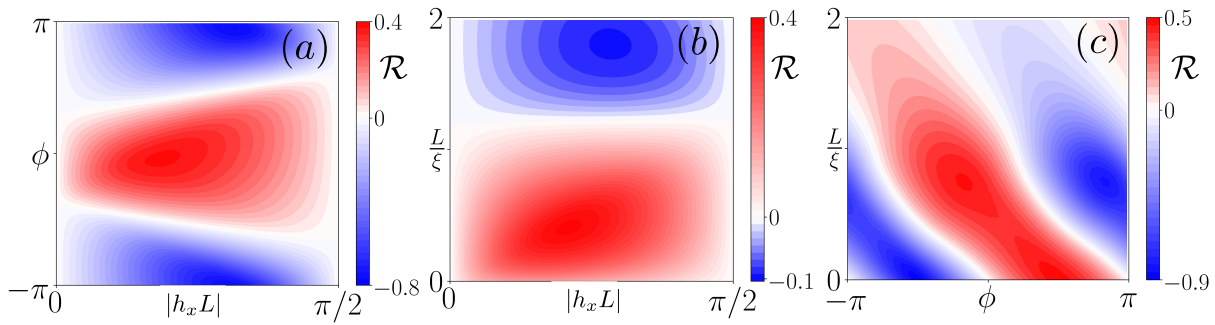

FIG. 3. Rectification coefficient ($\mathcal{R}$) in the (a) $h_xL - \phi$ plane for $L/\xi = 0.5$, (b) $h_xL - L/\xi$ plane for $\phi = 0.16\pi$, and (c) $\phi - L/\xi$ plane for $|h_xL| = \pi/4$. The rest of the parameters are the same as in Fig. 2.

## IV. RECTIFICATION COEFFICIENT

With the understanding of the behaviors of the thermal conductances, we now quantify the rectification by our WJJ thermal diode by defining the rectification coefficient ($\mathcal{R}$) as [89, 93],

$$\mathcal{R} = \frac{|\kappa_F| - |\kappa_R|}{|\kappa_F|} \quad (8)$$

where the positive (negative) values of $\mathcal{R}$ indicate a higher forward (reverse) thermal current compared to the reverse (forward) one and $\mathcal{R} = 0$ indicates no rectification. One can also define it by taking the ratio of the forward to the reverse thermal current and the sign changing phenomena will then be described by comparing the values with respect to 1. Thus, the behavior of the rectification coefficient will not be affected qualitatively. We emphasize here that this rectification coefficient is different from the efficiency of a device.

From Fig. 3(a), it is intriguing that the sign and the magnitude of the rectification coefficient is tunable by both the superconducting phase and Zeeman field externally. For a particular junction size and Zeeman field, the sign of the rectification coefficient can be tuned from the positive to the negative or vice versa by tuning the phase difference. It is true for both positive and the negative phase differences. Note that, it is possible to achieve TDE for $\phi = 0$ indicating that finite superconducting phase difference is not necessary for the TDE. On the other hand, a finite Zeeman field is essential to achieve TDE in our WJJ. For minimal values of $h_x$, the TDE is very low practically. The coefficient can be increased by tuning the Zeeman field. For a fixed value of the phase, the sign of $\mathcal{R}$ is mostly fixed. We find that the most efficient TDE condition $\mathcal{R} = -0.8$ can be achieved when $|h_xL| \sim \pi/4$. This observation holds quantitative significance. On the other hand, $\mathcal{R}$ reaches a notably high value $\sim -0.8$, indicating higher reverse quasi-particle current than the forward current, when $\phi \sim \pm\pi$ and $|h_xL| \approx \pi/4$. However the scenario changes when the junction size is different as we see in Fig. 3(b). For a fixed superconducting phase, we find two clearly separated regimes for the positive and negative sign of the rectification coefficient corresponding to the short ($L < \xi$) and long junction limit ($L > \xi$), respectively. However, the rectification is higher in the short junction limit and we find highest value of the rectification $\mathcal{R} \sim 40\%$ at $\phi = 0.16\pi$. The magnitude can be controlled by external Zeeman field as seen in Fig. 3(b).

The picture becomes complicated when we investigate the behavior of $\mathcal{R}$ in the $L/\xi - \phi$ plane as shown in Fig. 3(c). Now it is confirmed that for a particular junction size, the sign of the rectification coefficient can be tuned by tuning the superconducting phase. Explicitly, our WJJ based TDE is more efficient for this particular magnetic field when the junction size is small. For a short junction limit ($L < \xi$), we achieve the rectification coefficient as high as 90% just by tuning the phase. Here, we choose the value of $h_xL$ from Fig. 3(a) where the $\mathcal{R}$ is higher.

## V. SUMMARY AND CONCLUSION

We have theoretically shown TDE where the thermal currents are carried by the quasiparticles, with energy greater than the zero-temperature superconducting gap ($\epsilon > \Delta_0$), across a JJ made of an ISB WSM sandwiched between two WSCs. We have applied a Zeeman field ($h_x$) in the normal WSM region and maintained a temperature gradient across the junction. The Zeeman field allows a shifting of the Weyl nodes along the direction decided by their chiralities. This results in an asymmetry between the forward and reverse quasiparticle currents describing a TDE. We assume that, the Zeeman field is coupled only with the spin degree of freedom. The Zeeman energy is small compared to the energy scale associated with the separation between the Weyl nodes. It is justified for low magnetic field. We have achieved the thermal rectification as high as 90% in the short junction limit ($L \ll \xi$) leading towards an ideal diode effect. We have derived analytical expressions for the quasiparticles transmission for both long and short junctions to explain our numerical results. Most importantly, we have shown that the sign and magnitude of the rectification coefficient can be controlled externally by tuning the magnetic field and the superconducting phase. This tunability enhances the potential of our work significantly.

Throughout the manuscript, we have considered linear regime only because of the low thermal gradient applied across the junction. We have also neglected the phonon effect and considered only the electronic thermal conductance because of the low temperature gradient. In the ISB WSCs of our setup, we have considered spin-singlet pairing between Weyl nodes of same chirality since pairings between Weyl nodes of opposite chirality will be blocked due to the change in the chirality similar to other Weyl materials [94, 95]. We have neglected the inter-Weyl scatterings throughout our study since there is no long-range disorder in our system, and inter-Weyl scatterings are important in the presence of long-range disorder [96–98].

Noteworthy to mention, usage of Weyl materials for our TDE is justified as it is proved to be potential host for both intrinsic and proximity-induced superconductivity in the literature [90, 91, 94, 95, 97, 99–103] and also from the perspective of the current development of the field of thermal transport in superconducting junctions [76–88]. We have theoretically shown the TDE in a minimal model of WJJ. Our work on the TDE based on the quasiparticle thermal currents in 3D WJJ invites further study on the TDE based on superconducting junctions for the potential applications in designing various thermal devices. The effects of the inter-Weyl scatterings are planned to be communicated separately in our future work.

### ACKNOWLEDGMENTS

We acknowledge Arijit Saha for stimulating discussions and support throughout the project. We thank Amartya Pal for helpful discussions. P. D. acknowledges Department of Space (DoS), India for the entire support at PRL and also Department of Science and Technology (DST), India for the financial support through SERB Start-up Research Grant (File no. SRG/2022/001121) and the hospitality at Institute of Physics, Bhubaneswar, India where this work was initiated.

### Appendix A: Derivation of the eigenspinors

In this section, we show the derivation of Eq. (2) starting from the Eq. (1) of the main text as,

$$H(\mathbf{k}) = \begin{pmatrix} k_0^2 - |\vec{k}|^2 & 0 & k_x - ik_y & i\alpha k_y - \beta \\ 0 & k_0^2 - |\vec{k}|^2 & -i\alpha k_y + \beta & -k_x - ik_y \\ k_x + ik_y & i\alpha k_y + \beta & -(k_0^2 - |\vec{k}|^2) & 0 \\ -i\alpha k_y - \beta & -k_x + ik_y & 0 & -(k_0^2 - |\vec{k}|^2) \end{pmatrix}$$

We can rewrite the Hamiltonian in the second quantized notation as,

$$\begin{aligned}\mathcal{H} &= \sum_k \psi_{\vec{k}}^\dagger H(\vec{k}) \psi_{\vec{k}} \\ &= \sum_k \Big[ (k_0^2 - |\vec{k}|^2) c_{A\uparrow k}^\dagger c_{A\uparrow k} + (k_x + ik_y) c_{B\uparrow k}^\dagger c_{A\uparrow k} + (\beta - i\alpha k_y) c_{B\downarrow k}^\dagger c_{A\uparrow k} + (k_0^2 - |\vec{k}|^2) c_{A\downarrow k}^\dagger c_{A\downarrow k} + (\beta + i\alpha k_y) c_{B\uparrow k}^\dagger c_{A\downarrow k} \\ &\quad + (-k_x + ik_y) c_{B\downarrow k}^\dagger c_{A\downarrow k} + (k_x - ik_y) c_{A\uparrow k}^\dagger c_{B\uparrow k} + (\beta - i\alpha k_y) c_{A\downarrow k}^\dagger c_{B\uparrow k} - (k_0^2 - |\vec{k}|^2) c_{B\uparrow k}^\dagger c_{B\uparrow k} + (-\beta + i\alpha k_y) c_{A\uparrow k}^\dagger c_{B\downarrow k} \\ &\quad + (-k_x - ik_y) c_{A\downarrow k}^\dagger c_{B\downarrow k} - (k_0^2 - |\vec{k}|^2) c_{B\downarrow k}^\dagger c_{B\downarrow k} \Big]. \quad \text{(A1)} \end{aligned}$$

We can rewrite Eq. (A1) in terms of the new basis defined in the main text as,

$$\mathcal{H} = \sum_k \Big[ (k_0^2 - |\vec{k}|^2) c_{\uparrow k}^{A\,\dagger} c_{\uparrow k}^A + (k_0^2 - |\vec{k}|^2) c_{\downarrow k}^{A\,\dagger} c_{\downarrow k}^A + ik_y c_{\uparrow k}^{B\,\dagger} c_{\uparrow k}^A + (k_x - \beta - i\alpha k_y) c_{\uparrow k}^{B\,\dagger} c_{\downarrow k}^A + (k_x + \beta + i\alpha k_y) c_{\downarrow k}^{B\,\dagger} c_{\uparrow k}^A$$
$$+ ik_y c_{\downarrow k}^{B\,\dagger} c_{\downarrow k}^A - ik_y c_{\uparrow k}^{A\,\dagger} c_{\uparrow k}^B + (k_x + \beta - i\alpha k_y) c_{\uparrow k}^{A\,\dagger} c_{\downarrow k}^B + (k_x - \beta + i\alpha k_y) c_{\downarrow k}^{A\,\dagger} c_{\uparrow k}^B - ik_y c_{\downarrow k}^{A\,\dagger} c_{\downarrow k}^B - (k_0^2 - |\vec{k}|^2) c_{\uparrow k}^{B\,\dagger} c_{\uparrow k}^B$$
$$- (k_0^2 - |\vec{k}|^2) c_{\downarrow k}^{B\,\dagger} c_{\downarrow k}^B \Big]. \quad \text{(A2)}$$

Redefining the basis around the Weyl nodes $P_{i=1,2,3,4}$ as,

$$\chi_{1,\vec{k}} = \chi_{3,\vec{k}} = (c_{\uparrow,\vec{k}}^B, c_{\downarrow,\vec{k}}^A)^T \text{ and } \chi_{2,\vec{k}} = \chi_{4,\vec{k}} = (c_{\uparrow,\vec{k}}^A, c_{\downarrow,\vec{k}}^B)^T,$$

the Hamiltonian can be Taylor expanded around the Weyl nodes as

$$\mathcal{H}_w = \sum_k{}' \Big[ \chi_{1,\vec{k}}^\dagger H_1(k) \chi_{1,\vec{k}} + \chi_{2,\vec{k}}^\dagger H_2(k) \chi_{2,\vec{k}} + \chi_{3,\vec{k}}^\dagger H_3(k) \chi_{3,\vec{k}} + \chi_{4,\vec{k}}^\dagger H_4(k) \chi_{4,\vec{k}} \Big] \quad \text{(A3)}$$

where

$$H_1(k) = (k_x - \beta)\sigma_x + \alpha k_y \sigma_y + \Big[ 2\sqrt{k_0^2 - \beta^2}\Big(k_z - \sqrt{k_0^2 - \beta^2}\Big) + 2\beta(k_x - \beta) \Big]\sigma_z,$$
$$H_2(k) = (k_x + \beta)\sigma_x + \alpha k_y \sigma_y + \Big[ 2\sqrt{k_0^2 - \beta^2}\Big(k_z + \sqrt{k_0^2 - \beta^2}\Big) + 2\beta(k_x + \beta) \Big]\sigma_z,$$
$$H_3(k) = (k_x - \beta)\sigma_x + \alpha k_y \sigma_y + \Big[ -2\sqrt{k_0^2 - \beta^2}\Big(k_z + \sqrt{k_0^2 - \beta^2}\Big) + 2\beta(k_x - \beta) \Big]\sigma_z,$$
$$H_4(k) = (k_x + \beta)\sigma_x + \alpha k_y \sigma_y + \Big[ -2\sqrt{k_0^2 - \beta^2}\Big(k_z - \sqrt{k_0^2 - \beta^2}\Big) + 2\beta(k_x + \beta) \Big]\sigma_z. \quad \text{(A4)}$$

We have neglected the term $2\beta(k_x \pm \beta)\sigma_z$, where the $k_x$ is the momentum parallel to the interface, for the simplicity of the model since our idea was to show the possibility of diode effect using a minimal model for ISB Weyl junction.

Now, we present the eigenspinors in the WSC and WSM region in the presence of an external Zeeman field. We consider the quasi 1D transport along $z$ direction as shown in Fig. 1 in the main text. The eigenspinors of the WSM region can be written as

$$\Psi_{\vec{e}}(z) = \big(\cos\alpha_e, e^{i\theta_k}\sin\alpha_e, 0, 0\big)^T e^{ik_e^+ z},$$
$$\Psi_{\overleftarrow{e}}(z) = \big(e^{-i\theta_k}\sin\alpha_e, \cos\alpha_e, 0, 0\big)^T e^{ik_e^- z},$$
$$\Psi_{\vec{h}}(z) = \big(0, 0, -e^{-i\theta_k}\sin\alpha_h, \cos\alpha_h\big)^T e^{ik_h^+ z},$$
$$\Psi_{\overleftarrow{h}}(z) = \big(0, 0, \cos\alpha_h, -e^{i\theta_k}\sin\alpha_h\big)^T e^{ik_h^- z}, \quad \text{(A5)}$$

where $k_e^\pm = -h_x \pm \sqrt{(E+\mu_N)^2 - k_{||}^2}$, $k_h^\pm = h_x \pm \sqrt{(E-\mu_N)^2 - k_{||}^2}$ are the momenta of the electron and hole respectively, $\alpha_{(e,h)} = \tan^{-1}\big(k_{||}/\sqrt{(E\pm\mu_N)^2 - k_{||}^2}\big)/2$ are the incident angles of the incoming electron and hole respectively and $\theta_k = \tan^{-1}(k_y/k_x)$.

Similarly, the wave functions in the WSC region can be written as,

$$\Psi_{\overrightarrow{eq}}(z) = \left(e^{i\beta}\cos\tilde{\alpha}_e, e^{i\beta}e^{i\theta_k}\sin\tilde{\alpha}_e, e^{-i\Phi_l}\cos\tilde{\alpha}_e, e^{-i\Phi_l}e^{i\theta_k}\sin\tilde{\alpha}_e\right)^T e^{iq_e z},$$

$$\Psi_{\overleftarrow{eq}}(z) = \left(e^{i\beta}e^{-i\theta_k}\sin\tilde{\alpha}_e, e^{i\beta}\cos\tilde{\alpha}_e, e^{-i\Phi_l}e^{-i\theta_k}\sin\tilde{\alpha}_e, e^{-i\Phi_l}\cos\tilde{\alpha}_e\right)^T e^{-iq_e z},$$

$$\Psi_{\overrightarrow{hq}}(z) = \left(e^{i\Phi_l}e^{-i\theta_k}\cos\tilde{\alpha}_h, e^{i\Phi_l}\sin\tilde{\alpha}_h, e^{i\beta}e^{-i\theta_k}\cos\tilde{\alpha}_h, e^{i\beta}\sin\tilde{\alpha}_h\right)^T e^{-iq_h z},$$

$$\Psi_{\overleftarrow{hq}}(z) = \left(e^{i\Phi_l}\sin\tilde{\alpha}_h, e^{i\Phi_l}e^{i\theta_k}\cos\tilde{\alpha}_h, e^{i\beta}\sin\tilde{\alpha}_h, e^{i\beta}e^{i\theta_k}\cos\tilde{\alpha}_h\right)^T e^{iq_h z}, \tag{A6}$$

where, $q_{(e,h)} = \sqrt{(\mu_S \pm \Omega)^2 - k_\parallel^2}$ and $\tilde{\alpha}_{(e,h)} = \tan^{-1}\left(k_\parallel/q_{(e,h)}\right)/2$ are the momenta and incident angles of the electron-like and hole-like quasiparticles, respectively, inside the superconducting region. For $E \leq \Delta$ (sub-gap regime), $\beta = \cos^{-1}(E/\Delta)$ and $\Omega = i\sqrt{\Delta^2 - E^2}$, while for $E > \Delta$ (supra-gap regime), $\beta = -i\cosh^{-1}(E/\Delta)$ and $\Omega = \sqrt{E^2 - \Delta^2}$. For simplicity, we assume $\Phi_L = -\Phi_R = -\phi/2$ throughout the work.

### Appendix B: Scattering matrix formalism

Next, we discuss the major steps of the scattering matrix formalism that we use to find the thermal currents by the quasiparticles driven by a temperature gradient between left and right WSC region. The scattering states for electron-like quasiparticles transmission along the forward direction (as sketched in Fig. 1 in the main text) can be written as,

$$\Psi_s^L(z) = \Psi_{\overrightarrow{eq}} + r_{ee}^F \Psi_{\overleftarrow{eq}} + r_{he}^F \Psi_{\overleftarrow{hq}},$$
$$\Psi^N(z) = a^F \Psi_{\overrightarrow{e}} + b^F \Psi_{\overleftarrow{e}} + c^F \Psi_{\overrightarrow{h}} + d^F \Psi_{\overleftarrow{h}},$$
$$\Psi_s^R(z) = t_{ee}^F \psi_{\overrightarrow{eq}} + t_{he}^F \psi_{\overrightarrow{hq}}, \tag{B1}$$

where $r_{ee}^F$ ($r_{he}^F$) represents the ordinary (Andreev) reflection coefficient for electron-like quasiparticles. Here, $a^F$, $b^F$, $c^F$, and $d^F$ are the scattering amplitudes of the quasiparticles within the WSM region. The coefficients $t_{ee}^F$ ($t_{he}^F$) represent the transmission coefficient of the electron-like quasiparticles as electron-like (hole-like) quasiparticles. The scattering amplitudes follow orthonormality condition,

$$\left|r_{ee}^F\right|^2 + \left|r_{he}^F\right|^2 + \left|t_{ee}^F\right|^2 + \left|t_{he}^F\right|^2 = 1. \tag{B2}$$

We derive the scattering coefficients for the quasiparticles moving along the forward direction via wavefunction matching condition at the two junctions ($z = 0$ and $z = L$) given by,

$$\Psi_s^L(z=0) = \Psi^N(z=0)$$
$$\Psi^N(z=L) = \Psi_s^R(z=L). \tag{B3}$$

We carry out a similar analysis to obtain the quasiparticle transmission probability along the reverse direction assuming the particles are incident from the right side when we apply the negative thermal gradient. We calculate both the forward and reverse transmission probability using this scattering matrix formalism to find the current flowing through the thermally-biased WJJ.

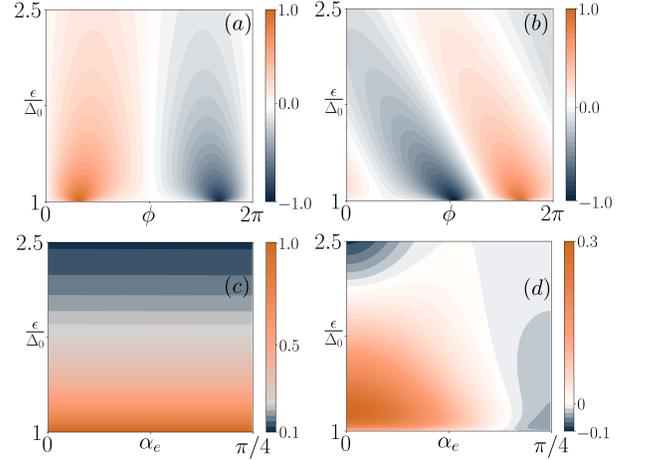

FIG. 4. $(T_F - T_R)$ in (a-b) $\epsilon/\Delta_0 - \phi$ plane for (a) $L = 0.01\xi$ and (b) $L = \xi$ with fixed angle of incidence $\alpha_e = 0.16\pi$, and (c-d) $\epsilon/\Delta_0 - \alpha_e$ plane for (a) $L = 0.01\xi$ and (b) $L = 0.5\xi$ keeping $\phi = 0.32\pi$. The rest of the parameters are: $\mu_N = 0.5\Delta_0$, $\mu_s = 100\Delta_0$, $T/T_c = 0.3$, and $|h_x L| = 0.5$.

### Appendix C: Asymmetry in the quasiparticles transmission probabilities

In this section, we present some additional results to show the mismatch in quasiparticle transmission probabilities ($T_F - T_R \neq 0$) which arises due to the shifting of the Weyl nodes in the opposite directions. This mismatch leads to the TDE in our WJJ shown in the main text.

In Fig. 4, we show density plots of the difference between the forward and reverse transmission probabilities of the quasiparticles ($T_F - T_R$) across the WJJ with energy greater than the superconducting gap ($\epsilon > \Delta_0$). Each panel of the figure corresponds to finite values of $T_F - T_R$ in the presence of the Zeeman field term $|h_x L|$, leading to the phenomenon of a diode effect shown in Fig. 3 in the main text. It is clear that there exists an asymmetry between the transmission probabilities along the opposite directions. For very short junction, $T_F - T_R$ can be tuned from the positive to the negative or the vice versa by tuning the superconducting phase difference. For the range $0 < \phi < \pi/2$, the difference is positive indicating that the forward transmission probability is higher than that in the reverse direction (see Fig. 4(a)). On the other hand, we get the opposite scenario i.e., higher re-verse transmission probability than the forward one when the phase is within the range $\pi/2 < \phi < \pi$. However, this phase-tunability changes with the change in the length of the junction. In Fig. 4(b), we show $T_\text{F} - T_\text{R}$ for relatively longer junction. The sign changing phenomenon by tuning the phase difference is still present. This asymmetry in the transmission probability results in TDE. To reveal the dependence of the transmission probability difference on the angle of incidence, we plot the same as a function of the angle of incidence and see that it is practically independent of the angle of incidence of the electron in the normal region ($\alpha_e$) for a short junction limit ($L \ll \xi$) as shown in Fig. 4(c). On the other hand, a clear $\alpha_e$ dependency is observed for an intermediate junction limit where $L = 0.5\xi$ (see Fig. 4(d)). For the latter plot, we choose $L/\xi$ where the difference $(T_\text{F} - T_\text{R})$ is relatively higher.

For the normal incidence of the quasiparticales and the short junction limit $(k(\epsilon)L \approx k(0)L)$, we have $\tilde{\alpha}_e, \tilde{\alpha}_h, \alpha_e, \alpha_h \approx 0$. Hence, the propagating wavevectors become in the normal region, $k^{\pm}_{e(h)} \approx \mp h_x$ and in the superconducting region, $q_{e,h} \sim 0$. Hence, the eigenspinors of the normal WSM region transfrom as,

$$\Psi_{\vec{e}}(z) = (1,0,0,0)^T e^{-ih_x z} ,$$
$$\Psi_{\cev{e}}(z) = (0,1,0,0)^T e^{ih_x z} ,$$
$$\Psi_{\vec{h}}(z) = (0,0,0,1)^T e^{ih_x z} ,$$
$$\Psi_{\cev{h}}(z) = (0,0,1,0)^T e^{-ih_x z} . \quad \text{(C1)}$$

Similarly, the wave functions in the WSC region can be written as,

$$\Psi_{\vec{eq}}(z) = \left(e^{i\beta}, 0, e^{-i\Phi_t}, 0\right)^T ,$$
$$\Psi_{\cev{eq}}(z) = \left(0, e^{i\beta}, 0, e^{-i\Phi_t}\right)^T ,$$
$$\Psi_{\vec{hq}}(z) = \left(e^{i\Phi_t} e^{-i\theta_k}, 0, e^{i\beta} e^{-i\theta_k}, 0\right)^T ,$$
$$\Psi_{\cev{hq}}(z) = \left(0, e^{i\Phi_t} e^{i\theta_k}, 0, e^{i\beta} e^{i\theta_k}\right)^T . \quad \text{(C2)}$$

Therefore, by solving Eq. (B1) and Eq. (B3), the quasiparticles transmission probability along the forward and reverse direction in the short ($L \ll \xi$) and long ($L \gg \xi$) junction limit can be written as Eqs. (5) and (6) in the main text, respectively.

---